%% file: sample-sigconf.tex
\documentclass[sigconf,10pt]{acmart}

\renewcommand\footnotetextcopyrightpermission[1]{} 
\usepackage{booktabs} 
\usepackage{graphicx}
\usepackage{array}
\usepackage{caption}
\usepackage{algorithm}
\usepackage{algorithmicx} 
\usepackage{algpseudocode}
\usepackage{amsmath}  
\usepackage{graphicx}
\usepackage{multirow}
\usepackage{enumitem}
\usepackage{subfig}

\usepackage{array}
\newcommand{\PreserveBackslash}[1]{\let\temp=\\#1\let\\=\temp}
\newcolumntype{C}[1]{>{\PreserveBackslash\centering}m{#1}}
\newcolumntype{R}[1]{>{\PreserveBackslash\raggedleft}m{#1}}
\newcolumntype{L}[1]{>{\PreserveBackslash\raggedright}m{#1}} 

\makeatletter
\newcommand\fs@spaceruled{\def\@fs@cfont{\bfseries}\let\@fs@capt\floatc@ruled
  \def\@fs@pre{\hrule height.8pt depth0pt \kern2pt}%
  \def\@fs@post{\kern2pt\hrule\relax\vspace{-10pt}}%
  \def\@fs@mid{\kern2pt\hrule\kern2pt}%
  \let\@fs@iftopcapt\iftrue}
\makeatother

\setcopyright{none}

\begin{document}
\title{Being-ahead: Benchmarking and Exploring Accelerators for Hardware-Efficient AI Deployment}

\author{\vspace{-10pt}\Large Xiaofan Zhang, Hanchen Ye, Deming Chen\\
        \large University of Illinois at Urbana-Champaign\\
        \large \textit{\{xiaofan3, hanchen8, dchen\}@illinois.edu}
}

\input{./sections/00-abstract.tex}

\maketitle
\input{./sections/01-introduction.tex}

\input{./sections/02-related.tex}

\input{./sections/03-overall_intro.tex}

\input{./sections/04-benchmark.tex}
\input{./sections/05-exploration.tex}

\input{./sections/06-results.tex}
\input{./sections/09-conclusion.tex}

\renewcommand*{\bibfont}{\footnotesize}

\bibliographystyle{unsrt}
\bibliography{sample-bibliography} 

\end{document}

%% file: sections/00-abstract.tex
\begin{abstract}

Customized hardware accelerators have been developed to provide improved performance and efficiency for DNN inference and training.
However, the existing hardware accelerators may not always be suitable for handling various DNN models as their architecture paradigms and configuration tradeoffs are highly application-specific. It is important to benchmark the accelerator candidates in the earliest stage to gather comprehensive performance metrics and locate the potential bottlenecks.
Further demands also emerge after benchmarking, which require adequate solutions to address the bottlenecks and improve the current designs for targeted workloads.
To achieve these goals, in this paper, we leverage an automation tool called DNNExplorer \cite{zhang2020dnnexplorer} for benchmarking customized DNN hardware accelerators and exploring novel accelerator designs with improved performance and efficiency. Key features include (1) direct support to popular machine learning frameworks for DNN workload analysis and accurate analytical models for fast accelerator benchmarking; (2) a novel accelerator design paradigm with high-dimensional design space support and fine-grained adjustability to overcome the existing design drawbacks; and (3) a design space exploration (DSE) engine to generate optimized accelerators by considering targeted AI workloads and available hardware resources. 
Results show that accelerators adopting the proposed novel paradigm can deliver up to 4.2$\times$ higher throughput (GOP/s) than the state-of-the-art pipeline design in \cite{zhang2018dnnbuilder} and up to 2.0$\times$ improved efficiency than the recently published generic design in \cite{ye2020hybrid}) given the same DNN model and resource budgets.
With DNNExplorer's benchmarking and exploration features, we can be ahead at building and optimizing customized AI accelerators and enable more efficient AI applications.

\end{abstract}

%% file: sections/01-introduction.tex
\vspace{-6pt}
\section{Introduction}


DNNs have been achieving great successes in facilitating a wide range of artificial intelligence (AI) applications. With the continuous improvement of neural network models, DNNs begin to adopt more sophisticated and efficient layer interconnections and deliver the state-of-the-art solutions for popular domains such as computer vision \cite{krizhevsky2012imagenet,simonyan2014very,szegedy2015going,he2016deep,redmon2016you,zoph2018learning,real2019regularized,zhang2020skynet} and natural language processing \cite{bengio2003neural, mikolov2010recurrent,bahdanau2014neural,wu2016google,gehring2016convolutional,vaswani2017attention}. 
Meanwhile, the impressive quality of results comes with the increasing compute and memory demands, which rely on building customized hardware accelerators to accommodate the DNN models and deliver high inference accuracy and satisfy inference speed, throughput, and energy efficiency. 

There are rich studies on customized DNN accelerators that take advantage of different hardware architectures for emerging AI applications, such as the neural processing unit \cite{esmaeilzadeh2012neural}, DianNao series \cite{chen2016diannao}, Eyeriss \cite{isscc_2016_chen_eyeriss}, and the tensor processing unit (TPU) \cite{jouppi2017datacenter}.
In addition, recently published literature also focuses on building customized accelerators on FPGAs for more flexible configurations, faster time-to-market, and improved latency and energy efficiency
\cite{qiu2016going,zhang2017high,zhang2017machine,chen2019cloud,zhang2018dnnbuilder,ye2020hybrid}.
By investigating these designs, we see two popular architecture paradigms: as the one uses customized implementations for DNN layers and forms a layer-wise pipeline architecture with dedicated pipeline stage handling each DNN layer \cite{li2016high,zhang2018dnnbuilder,wei2018tgpa}; while the other adopts a generic reusable architecture that all DNN layers are processed recurrently \cite{isscc_2016_chen_eyeriss,jouppi2017datacenter,zhang2017improving,hao2019fpga,ye2020hybrid}.

By following different architecture paradigms and design combinations, customized accelerators may present completely different performances when handling emerging DNN models, where more diverse layer configurations and deeper network structures are involved. 
For example, the first paradigm has an obvious flaw as it requires dedicated hardware instances for handling each pipeline stage. The more layers in the DNN models mean the fewer resources for each stage by considering the limited resource budget. The second paradigm has difficulty optimizing DNNs with various layers that exhibit vastly different arithmetic intensity because all layers share the generic architecture.  


It will be beneficial to have an efficient tool for benchmarking hardware accelerators with quantitative evaluations in the early stage of accelerator development. 
In this paper, we propose such a tool called DNNExplorer \cite{zhang2020dnnexplorer}. It enables an efficient accelerator design process through accelerator benchmarking and exploration features to accommodate demanding AI applications. Contributions of this work are summarized as follows:
(1) We propose an automation tool that provides seamless connections to popular machine learning frameworks (e.g., Caffe \cite{jia2014caffe} and PyTorch \cite{paszke2019pytorch}), where DNNs developed by these frameworks can be used to benchmark customized accelerators. It can accurately evaluate the performance of accelerators following both layer-based pipeline and generic reusable architecture.
%
(2) We introduce a novel architecture paradigm to overcome the respective drawbacks of the aforementioned popular paradigms. This unique paradigm enables high-dimensional design space as well as fine-grained adjustability and better scalability, which help deliver even better-customized accelerators.
(3) To efficiently explore the high-dimensional design space, a two-level design space exploration (DSE) engine is proposed to generate optimized accelerator configurations.
Results show that the proposed novel paradigm delivers up to 4.2$\times$ higher throughput than the pure pipeline accelerator \cite{zhang2018dnnbuilder} and up to 2.0$\times$ higher efficiency than the generic accelerator design \cite{ye2020hybrid}) when targeting the same DNN and hardware.

%% file: sections/02-related.tex
\vspace{-6pt}
\section{Related Work} 

With the diverse DNN workloads and various hardware accelerator designs, it is critical to understand how these different workloads perform on particular hardware configurations. Building performance benchmarks is one of the most effective approaches to distinguish competing accelerator designs. To achieve this goal, microbenchmarks are developed (e.g., DeepBench \cite{deepbench}) to measures the performance of basic DNN operations. 
Further efforts have been made to deliver system-level benchmarking. For example, TensorFlow Benchmarks collect popular image classification models to evaluate the throughput performance during DNN training \cite{tfbench}; and Training Benchmarks for DNNs (TBD) provide GPU evaluation when handling DNNs for computer vision and natural language processing \cite{zhu2018benchmarking}. To benchmark both hardware and software, DAWNBench is built to evaluate the performance of hardware and the accuracy of DNN models \cite{coleman2017dawnbench}.
Following the same idea, recently published benchmarks start leveraging more diverse tasks and developing more fair comparisons for different hardware setups. To achieve this goal, MLPerf is published through the joint efforts of academia and industry, which aims at providing unbiased evaluations of DNN training and inference performance \cite{mattson2020mlperf}.

In addition, researchers are keen on developing efficient performance modeling and hardware design methodologies to deliver desired customized accelerators for emerging AI applications. 
One of the popular research directions is building hardware accelerators from a higher abstraction level, such as behavioral level instead of register-transfer level to improve design efficiency \cite{rupnow2011study, liu2016high}. In \cite{zhang2017high}, a customized accelerator for image captioning is implemented on FPGA using high-level synthesis (HLS). Following the HLS-based design, more accelerators have been developed to meet the needs of various AI applications, such as accelerating image classification \cite{chen2019cloud,wei2019overcoming}, face recognition \cite{zhuge2018face,wang2020hardware}, object detection \cite{zhang2019skynet,hao2019fpga}, and language translation \cite{li2019implementing,li2020ftrans,qin2021}.

The other direction is to build automation design frameworks that provide systemic solutions to leverage the critical steps of DNN accelerator design, such as performance evaluation, architecture optimization, and fast prototyping.
A framework published in \cite{wei2017automated} introduces a systolic array based accelerator design along with an analytical model to estimate performance and resource utilization. In \cite{zhang2018caffeine}, a unified representation for convolutional (CONV) and fully-connected (FC) layers is proposed for accelerator modeling. To improve accelerator design and optimization, DNNBuilder is proposed to provide an end-to-end automation tool for building and prototyping high-quality accelerators \cite{zhang2018dnnbuilder}. It introduces a configurable fine-grained pipeline architecture and optimization algorithms to deliver real-time DNN inference even for resource-constraint hardware. In \cite{wang2018design}, accelerators generated by the proposed framework can also target extremely low bit-width DNNs where binary and ternary quantization schemes are adopted to replace the hardware-intensive floating-point multiplications by logical operations. Recently developed frameworks include more comprehensive hardware modeling methods, such as covering both spatial- and Winograd-CONV \cite{ye2020hybrid}, targeting accelerators for both FPGAs and ASICs \cite{xu2020autodnnchip}, and supporting multi-branch DNNs for virtual reality applications \cite{zhang2021f}.

Efficient approaches to benchmark and evaluate hardware accelerators also bring significant benefits to the emerging DNN-accelerator co-design strategies \cite{hao2019fpga, jiang2019accuracy,zhang2020skynet,li2020edd}. For example, authors in \cite{hao2019fpga} introduce a co-design method to simultaneously develop a hardware-efficient DNN model and high-performance customized accelerator given predefined applications and hardware constraints. One of the critical factors is the timely hardware performance feedback, which continuously provides design guidelines. Similarly, SkyNet is developed by fully considering the hardware constraints of both FPGAs and GPUs to deliver more efficient solutions to handle object detection and tracking \cite{zhang2020skynet}.

%% file: sections/03-overall_intro.tex
\vspace{-6pt}
\section{The Overall Flow of DNNExplorer}

\begin{figure}[t]
  \centering
  \includegraphics[width=0.9\columnwidth]{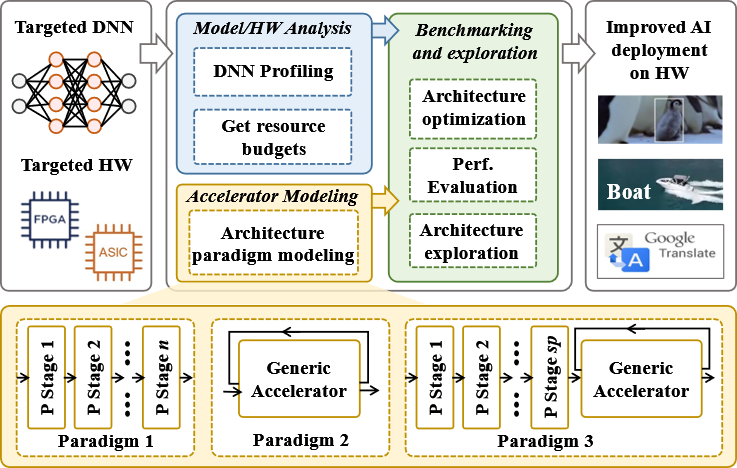}
  \vspace{-6pt}
   \caption{DNNExplorer introduces an complete flow for customized accelerator benchmarking and exploration to deliver improved AI hardware deployment.}
   \label{fig:dnnexplorer_flow}
\end{figure}

DNNExplorer is an automation framework that helps close the gap between fast DNN construction in software 
and slow AI hardware deployment. It can be fed in DNN models described by the high-abstraction network definition files (e.g., caffe prototxt files and PyTorch network forward functions) and deliver optimized customized accelerators by considering various user-defined constraints, such as resource budgets, accelerator architecture paradigms, and targeted hardware performance.  
As shown in Figure \ref{fig:dnnexplorer_flow}, DNNExplorer features three major steps as model/hardware analysis, accelerator modeling, and benchmarking and exploration. 

In step 1, DNN definition files are passed to DNNExplorer as inputs for model profiling. The layer-wise information is extracted, such as layer type, layer configuration, computation and memory demands, arithmetic intensity, quantization scheme, etc. The targeted hardware specifications are also entered to help setup resource budgets. For FPGA implementation, DNNExplorer captures three major resources as DSP, BRAM, and external memory bandwidth.

During step 2, DNNExplorer adopts three accelerator models corresponding to two popular paradigms (paradigm 1: the layer-based pipeline architecture \cite{li2016high,zhang2018dnnbuilder,wei2018tgpa} and paradigm 2: the generic reusable architecture \cite{zhang2017improving,hao2019fpga,ye2020hybrid}) and one proposed hybrid paradigm (paradigm 3). This step aims to adopt highly-accurate pre-built analytical models for hardware resource utilization and performance estimation. These models, along with the step 1 outputs, are then passed to step 3 for architecture optimization, benchmarking, and exploration. 

In step 3, optimization is performed to generate configuration guidelines. For accelerator designs following paradigm 1 or 2, respective optimization procedures (Section \ref{sec:accelerator_optim}) are activated to configure the accelerator and attempt to achieve maximum performance given resource constraints. DNNExplorer then starts accelerator benchmarking by evaluating the optimized accelerator using the targeted DNN model. 
For designs following the proposed paradigm 3, both layer-based pipeline architecture and generic reusable architecture are involved, creating more diverse configurations for achieving better performance. However, it also creates an enormous design space and is challenging to search for the most suitable configuration. An architecture exploration function is proposed to address these difficulties (Section \ref{sec:exploration}). This function contains a two-level automatic DSE engine following a divide-and-conquer strategy to determine task and resource partitioning schemes for both pipeline and generic architecture at the first level optimization and explore their respective optimal configuration with given resources at the second level optimization.
Eventually, DNNExplorer can effectively benchmark customized accelerators and explore novel architectures to deliver improved AI acceleration on given hardware.

%% file: sections/04-benchmark.tex
\vspace{-6pt}
\section{DNNExplorer for accelerator benchmarking}

The first two paradigms shown in Figure \ref{fig:dnnexplorer_flow} can be benchmarked after respective architecture optimizations. We adopt optimization strategies published in DNNBuilder \cite{zhang2018dnnbuilder}, such as the fine-grained pipeline and the column-based cache scheme to ensure the pipeline architecture is fully optimized. Regarding the generic architecture, we follow the optimization designs from HybridDNN \cite{ye2020hybrid} to enable a hybrid CONV processing engine (for spatial and Winograd CONV) and multi-dataflow support covering input stationary (IS) and weight stationary (WS). 
For the third paradigm, we need one more step as architecture exploration by using the proposed two-level DSE engine to deliver optimized hardware configuration, which will be introduced in Section \ref{sec:exploration}

\subsection{Pipeline architecture overview}

\begin{figure}
    \centering
    \includegraphics[width=0.4\textwidth]{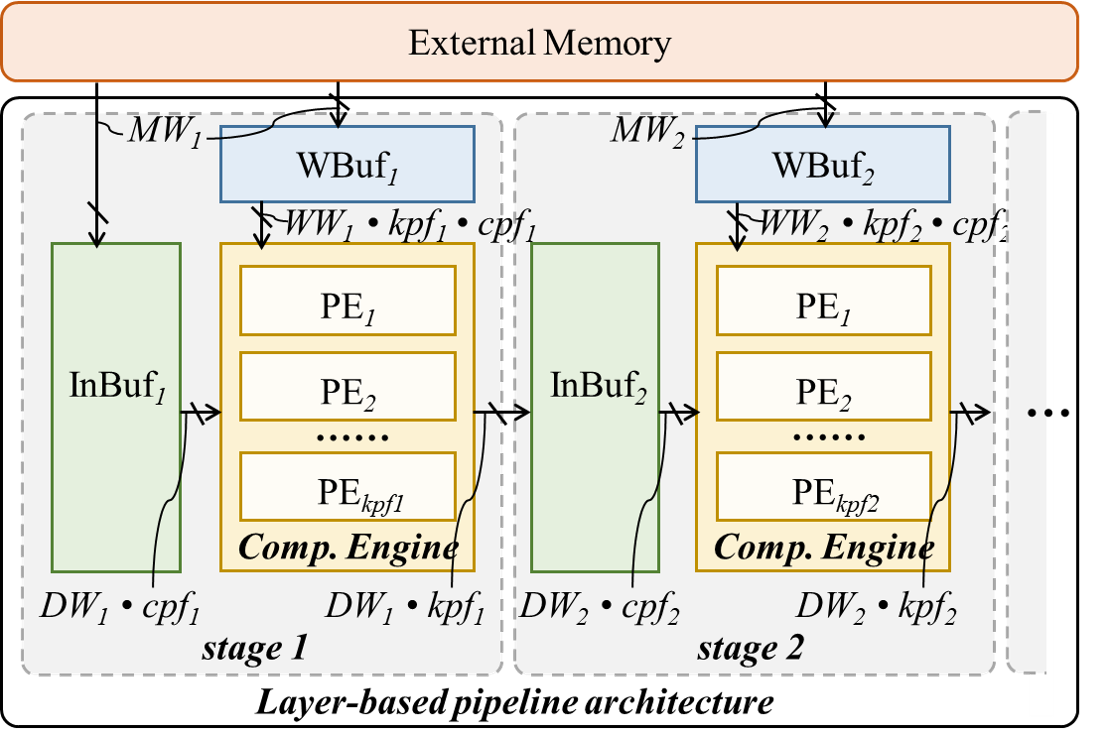}
    \vspace{-6pt}
    \caption{The layer-based pipeline architecture with dedicated pipeline stages for major DNN layers dominating computation and memory consumption.
    }
    \vspace{-12pt}
    \label{fig:pip_arch}
\end{figure}

As shown in Figure \ref{fig:pip_arch}, each stage is designed for accommodating major layers of the targeted DNN, such as CONV, Pooling (POOL), and fully-connected (FC) layers. Other layers, such as batch normalization (BN) and activation layers, are concatenated into the major ones and processed by the same pipeline stage.    
During implementation, three types of resources are required: the computation resources for building computation engines (CEs), the on-chip memory for implementing input and weight buffers, and the external memory for keeping DNN parameters. Configurable parameters are available in every stage, including channel parallelism factor ($CPF$), kernel parallelism factor ($KPF$), the input data bit-width (DW), the weight bit-width (WW), and the bit-width for external memory access (MW). Among these parameters, $CPF$ and $KPF$ are the unrolling factors along input and output dimensions, representing the number of input channels and the number of kernels being processed in parallel. $CPF$ and $KPF$ are calculated by the resource allocation algorithm, which will be introduced in Subsection \ref{sec:pip_optim}.

Following this paradigm, the Computation Engine (CE) is responsible for handling most of the computations. Inside each CE, there are processing elements (PEs) with two-dim parallelism ($CPF$ and $KPF$). More specifically, assuming stage 1 in Figure \ref{fig:pip_arch} works for the CONV layer, the CE in this stage contains $KPF_1$ PEs, and each PE is designed to handle one $CPF_1$-length vector multiplication in one clock cycle.
Once the DNN parameters are ready in weight buffer, we broadcast a $CPF_1$-length vector of input feature map to all $KPF_1$ PEs. After calculations, results ($KPF_1$-length vector) will be written to the input buffer of the next stage.

In addition, to ensure lower initial latency and efficient on-chip memory utilization of the pipeline architecture, we adopt the fine-grained layer-based pipeline and column-based cache scheme \cite{zhang2018dnnbuilder}. With these technologies, we no longer need to wait for the full completion of intermediate results before continuing to the next pipeline stage. 

\subsection{Generic architecture overview}
\label{sec:generic_arch}

\begin{figure}
    \centering
    \includegraphics[width=0.4\textwidth]{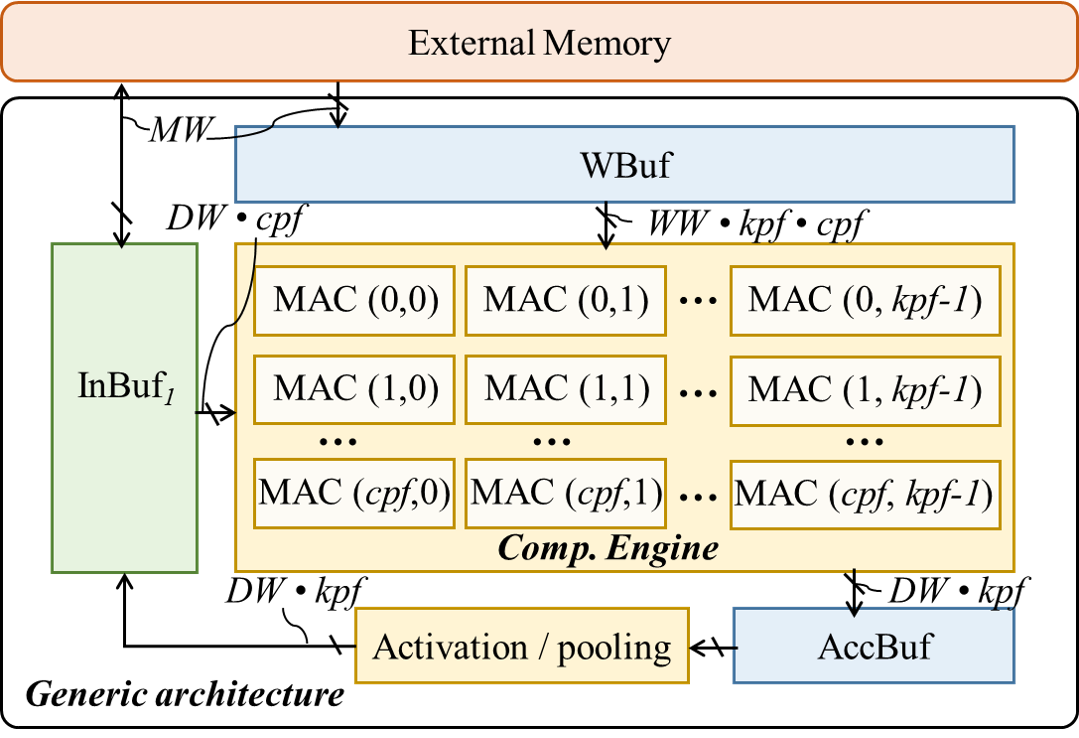}
    \vspace{-6pt}
    \caption{The reusable generic architecture adopted by DNNExplorer.}
    \vspace{-12pt}
    \label{fig:gen_arch}
\end{figure}

As shown in Figure \ref{fig:gen_arch}, the key component is a reusable MAC array, which is composed of $CPF_g \times KPF_g$ MAC units. Three on-chip buffers (feature map buffer, weight buffer, and accumulation buffer) are allocated for pumping data to the MAC array and storing the intermediate results. Similar to the pipeline architecture, $DW_g$, $WW_g$, and $MW_g$ denote the bit-width of input data, weight, and external memory bus. A functional module is instantiated between the accumulation buffer and feature map buffer for handling pooling and activation layers. In each clock cycle, the MAC array consumes a $CPF_g$-length vector of input feature map and a $CPF_g \times KPF_g$ matrix of DNN parameters and calculates one general matrix-vector multiplication (GEMV). The intermediate results are stored and accumulated in the accumulation buffer until all associated calculations are completed. 

In paradigm 2, the buffer allocation strategy becomes one of the most important design choices. Allocating a large feature map buffer will effectively reduce the frequency of loading/saving intermediate feature maps from/to the external memory. Allocating a large weight buffer will provide more freedom for the computation scheduling, potentially benefiting the performance. DNNExplorer supports a flexible on-chip buffer allocation strategy to enable a comprehensive exploration in the trade-off design space.
In addition, there are two supported data reuse strategies as Input Stationary (IS) and Weight Stationary (WS). For IS, input feature maps are partitioned into multiple groups along the height dimension and transferred from the external memory to the on-chip buffer by groups. One group of input feature maps will be kept on-chip and reused until all associated computations are completed. For WS, weights are partitioned along the output channel dimension, and all input feature maps are streamed in for each group of weights. 

\subsection{Accelerator modeling and optimization}
\label{sec:accelerator_optim}
One of the important features to enable accelerator benchmarking is the fast and accurate estimation of hardware performance and resource utilization. In this work, we adopt highly-accurate analytical models to generate estimated performance and resource utilization based on the DNN layer-wise information and optimized hardware configurations.
%
Our tool supports most of the commonly-used DNN layers (e.g., CONV, POOL, and FC layers) and can be extended to support more emerging DNN layers. In this section, we take a $n$-layer convolutional neural network as an example and assume the $i$-th CONV layer with a 3-dim input feature $D_i$ (size $H_i \times W_i$ with $CHin_i$ channels) and a 4-dim kernel $G_i$ (size $R_i \times S_i$ with $CHout_i$ output and $CHin_i$ input channels). For evaluating the performance, we assume $FREQ$ as the working clock frequency.

\subsubsection{The pipeline architecture}
\label{sec:pip_optim}
When configuring the pipeline architecture, we follow a particular resource allocation guideline for maximizing the overall hardware performance. Assuming the computation demand of DNN layer $i$ is $C_i$, the increase of allocated resource $R_i$ for layer $i$ results in a proportional increase in parallelism (meaning larger $CPF_i$ and $KPF_i$ are used) and eventually lowers the latency $L_i$ for that layer. By changing these parallelism factors, we can balance the DNN workloads by coordinating every pipeline stage and achieve the maximum throughput once the workload for each stage is well-balanced. The throughput performance can be described by Equation \ref{eq:pip_tp}, where $Batch$ indicates the batch size and the latency for layer $i$ can be shown in Equation \ref{eq:pip_lat}.

\begin{equation}
\label{eq:pip_tp}
\small
    Throughput = \frac{Batch}{max(L_1, L_2, ..., L_n)} 
\end{equation}

\begin{equation}
\label{eq:pip_lat}
\small
    L_i = \frac{H_i \times W_i \times R_i \times S_i \times CHin_i \times CHout_i}{CPF_i \times KPF_i \times FREQ}
\end{equation}

To achieve the optimized performance, we follow strategies proposed in \cite{zhang2018dnnbuilder} to complete computation and memory resource allocation. In Algorithm \ref{alg:pip_comp}, we allocate the computation resource to balance the latency of each pipeline stage. Since the parallelism factors must be the power of 2, we further fine-tune the allocation scheme and fills up the gap between the actual and the theoretical values. Resource allocated for layer $i$ is represented as $R_i$.

Then, we adopt Algorithm \ref{alg:pip_mem} to allocate memory bandwidth given the constraints of total bandwidth $BW_{total}$ and total amount of on-chip memory $mem_{total}$ for input buffers.
We initialize $Col_i=1$ (caching one column of the input feature map according to the column-based cache scheme) and the input buffer is implemented by a dual-port RAM with the width of read/write port $width^{rd}_{i}$, $width^{wr}_{i}$, and the depth of read port $depth^{rd}_{i}$ in $i$-th layer.
Algorithm \ref{alg:pip_mem} first satisfies the bandwidth demands of allocated computation resources (line 5). If the required memory bandwidth exceeds the total available bandwidth, we need bandwidth adjustment (starting from line 6). 
In this algorithm, $H^{in}_i$, $H^{out}_i$ and $Stride_i$ represent the height of input and output feature maps and the stride in layer $i$. $CHin_i$ and $CHout_i$ represent the number of channels of the input  and output feature map in layer $i$, respectively. $Col_i$ represents the number of cached columns in layer $i$, which relates to the kernel reuse behavior and the consumption of on-chip memory $mem_i$.

\begin{algorithm}[t] 
\renewcommand\baselinestretch{0.7}\selectfont
\small
\caption{Computation resource allocation (paradigm 1)\vspace{-2pt}}  
  \begin{algorithmic}[1]
    \State  Set available computation resource: $R_{total}$
    \State  Computation resource for layer $i$:  $R_i=\frac{C_i}{C_{total}} \times R_{total}$ 
    \State  Initialize allocated resource for $i$-th layer (parallelism factor): \\ $R_i = 2^{\lfloor\log _2 R_i\rfloor}$
    \State  \textbf{while} $\sum\nolimits_{i=1}^{n} R_i \leq R_{total}$
    \State  \quad Select layer $j$ with maximum $\frac{C_j}{R_j}$
    \State  \quad \textbf{if} $\sum\nolimits_{i=1}^{n} R_i + 2 \times R_j \leq R_{total}$
    \State  \quad \quad $R_j = 2 \times R_j$ //double the resource for layer $j$
    \State  \quad \textbf{else} break
    \State  $R_i = CPF_i \times KPF_i$
  \end{algorithmic}
  \vspace{-2pt}
  \label{alg:pip_comp}
\end{algorithm}

\begin{algorithm}[t] 
\renewcommand\baselinestretch{0.8}\selectfont
\small
\caption{Bandwidth resource allocation (paradigm 1)\vspace{-2pt}}  
  \begin{algorithmic}[1]
    \State  Set available memory bandwidth: $BW_{total}$
    \State  Set available on-chip memory for input buffers: $mem_{total}$
    \State  Set single DSP's bandwidth usage: $BW_R$
    \State  Initialize $Col_i$ = 1; size of input buffer (e.g. $width^{rd}_{i}$, $depth^{rd}_{i}$, and $width^{wr}_{i}$ according to $PF_i = KPF_i \times CPF_i$)
    \State  Allocate bandwidth $BW_i$ for layer $i$ to best satisfy its $R_i$ demand: \quad $BW_i=\frac{PF_i \times BW_R}{H^{out}_i \times Col_i}$
    \State  \textbf{while} $\sum\nolimits_{i=1}^{n} BW_i \geq BW_{total}$ //CONV layer bandwidth overuse 
    \State  \quad Select layer $i$ in CONV layer with maximum $BW_i$  
    \State  \quad $depth^{rd}_{i} += \frac{H^{in}_i \times CHin_i \times Stride_i}{CPF_i}, depth^{rd}_{i+1} += \frac{H^{out}_i \times CHout_i}{CPF_{i+1}}$
    \State  \quad \textbf{if} $\sum\nolimits_{i=1}^{n} f(width^{rd}_{i}, depth^{rd}_{i}, width^{wr}_{i}) \leq mem_{total}$ 
    \State  \quad \quad $Col_i = Col_i + 1$ //Cache one more column
    \State  \quad \quad $BW_i = BW_i \times \frac{Col_i-1}{Col_i}$ 
    \State  \quad \textbf{else}  //restore if not enough memory
    \State  \quad $depth^{rd}_{i} -= \frac{H^{in}_i \times CHin_i \times Stride_i}{CPF_i}, depth^{rd}_{i+1} -= \frac{H^{out}_i \times CHout_i}{CPF_{i+1}}$, break
  \end{algorithmic}
    \label{alg:pip_mem}
\end{algorithm}

\begin{figure}[t]
    \centering
    \includegraphics[width=0.51\textwidth]{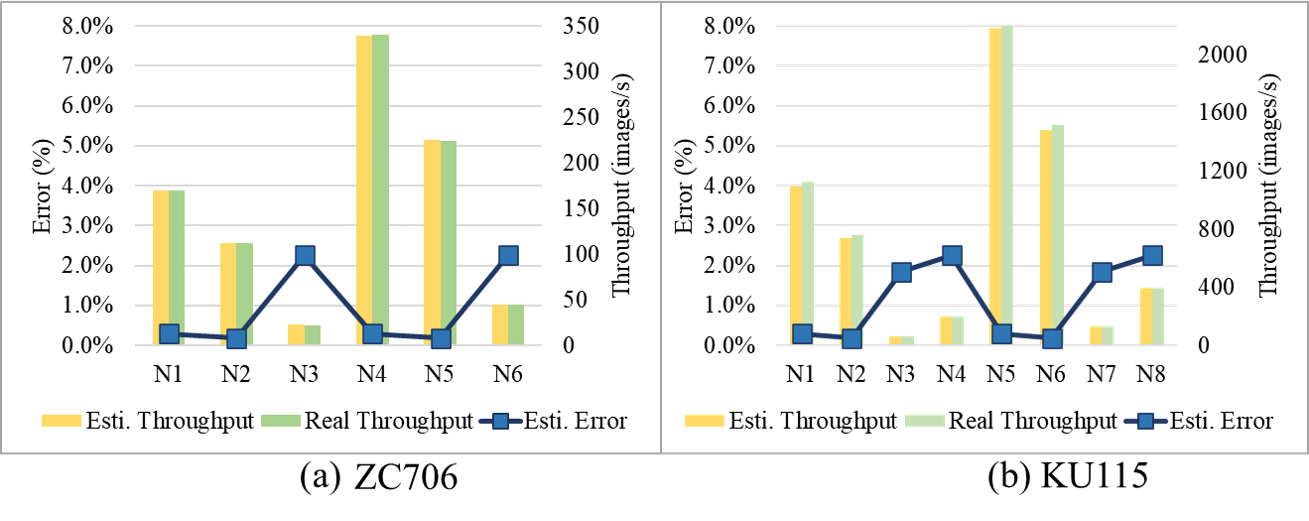}
    \vspace{-18pt}
    \caption{Performance estimation errors of the pipeline architecture. (a) N1$\sim$N3 represent AlexNet, ZF, and YOLO (16-bit quantization) while N4$\sim$N6 represent the same group of networks using 8-bit quantization. In (b), N1$\sim$N4 represent AlexNet, ZF, VGG16, and YOLO with 16-bit quantization while the rest networks using 8-bit quantization.}
    \vspace{-10pt}
    \label{fig:esti_error_pip}
\end{figure}

To demonstrate the accuracy of our proposed analytical models, we compare the estimated throughput performance of the customized accelerators to their board-level implementation using FPGAs. 
After adopting the optimization strategies mentioned in this subsection, we evaluate accelerators on an embedded FPGA (Xilinx ZC706) and a mid-range FPGA (Xilinx KU115) and show the estimation errors in Figure \ref{fig:esti_error_pip}.
The average error between the estimated and the board-level performance is 1.15\%.
%
The reason for achieving such a low error rate comes from the high degree of determinism. With the dedicated compute and control units instantiated on hardware, the execution time of running targeted workloads is deterministic, which helps design highly accurate analytical models.

\subsubsection{The generic architecture}
The computation latency of a DNN layer running on the generic architecture is determined by the parallel factor $CPF_g$ and $KPF_g$ as:
\begin{equation}
\small
    L_{comp} = \frac{H \times W \times R \times S \times CHin \times CHout}{CPF_g \times KPF_g \times FREQ}
\end{equation}

To precisely estimate the memory accessing latency, we divide the external memory bandwidth $BW$ into three portions: $BW_{w}$, $BW_{ifm}$, and $BW_{ofm}$, corresponding to the bandwidth for weight loading, feature map swapping in and out, respectively. The memory accessing latency $L_{w}$, $L_{ifm}$, and $L_{ofm}$ can be calculated as:
\begin{equation}
\small
    L_{w} = \frac{R \times S \times CHin \times CHout \times WW_g}{BW_{w}}
\end{equation}
\begin{equation}
\small
    L_{ifm} = \frac{H \times W \times CHin \times DW_g}{BW_{ifm}}
\vspace{-1pt}
\end{equation}
\begin{equation}
\small
    L_{ofm} = \frac{H \times W \times CHout \times DW_g}{BW_{ofm}}
\end{equation}

For the IS strategy, assuming the capacity (in bits) of the accumulation buffer is $CAP_{abuff}$, the input and output feature maps will be partitioned into $G_{fm}$ groups where:

\vspace{-2pt}
\begin{equation}
\small
    \label{equ:g_fmap}
    G_{fm} = \frac{H \times W \times CHout \times DW_g}{CAP_{abuff} / 2} .
\end{equation}

In Equation \ref{equ:g_fmap}, $CAP_{abuff}$ is divided by a factor of 2 because the using of ping-pong buffers to avoid data pollution. Since the generic architecture needs to fetch weights $G_{fm}$ times for IS, the overall latency of this CONV layer are:

\vspace{-2pt}
\begin{equation}
\small
    \label{equ:l_all_acc1}
    L_{is} = max(L_{comp}, L_{w} \times G_{fm}, L_{ifm}, L_{ofm})
\end{equation}


\begin{algorithm}[t] 
\renewcommand\baselinestretch{0.8}\selectfont
\small
\caption{Generic architecture DSE (paradigm 2)\vspace{-2pt}}  
    \begin{algorithmic}[1]
    \State \textbf{STEP1:} Search for hardware parameters choices
    \State Set the available resource number as $N_{dsp}$, $N_{bram}$, and $N_{lut}$
    \State Initialize $CPF_g$, $KPF_g$, and $hwParamsList$
    \State \textbf{while} (true):
    \State \quad Calculate $n_{dsp}$, $n_{bram}$, and $n_{lut}$ with resource model
    \State \quad \textbf{if} ($n_{dsp}$ > $N_{dsp}$ \textbf{or} $n_{bram}$ > $N_{bram}$ \textbf{or} $n_{lut}$ > $N_{lut}$) \textbf{break}
    \State \quad $hwParamsList$.append([$CPF_g$, $KPF_g$])
    \State \quad Update $CPF_g$ and $KPF_g$
    \State

    \State \textbf{STEP2:} Search for the optimal DNN mapping scheme for each hardware parameters choices
    \State Initialize $dataflowsList$ and $latencyList$
    \State \textbf{for} $hwParams$ \textbf{in} $hwParamsList$:
    \State \quad Initialize $dataflows$ and $latency$
    \State \quad \textbf{for} all $layer$:
    \State \quad \quad Find the best $dataflow$ and corresponding $layerLatency$
    \State \quad \quad $dataflows$.append($dataflow$)
    \State \quad \quad $latency$ += $layerLatency$
    \State \quad $dataflowsList$.append($dataflows$) 
    \State \quad $latencyList$.append($latency$)
    \State
    
    \State \textbf{STEP3:} Search for final solution with the lowest latency
    \State $optLatency$, $idx$ = \textbf{min}($latencyList$), \textbf{argmin}($latencyList$)
    \State \textbf{return} $hwParamsList$[$idx$], $dataflowsList$[$idx$], $optLatency$
    \end{algorithmic}
    \label{alg:gen_dse}
\end{algorithm}

Fro the WS strategy, we partition weights into $G_{w}$ groups along the output channel dimension $CHout$. If we use $CAP_{wbuff}$ to represent the capacity of weight buffer, $G_{w}$ can be calculated as:
\vspace{-1pt}
\begin{equation}
\small
    G_{w} = \frac{R \times S \times CHin \times CHout \times WW_g}{CAP_{wbuff}/2}
\vspace{-2pt}
\end{equation}

WS strategy keeps weights on-chip, and for each group of weights, all input feature maps need to be loaded before any computations. Since there are total $G_{w}$ groups of weights to be calculated, the overall latency is:

\vspace{-2pt}
\begin{equation}
\small
    \label{equ:l_all_acc2}
    L_{ws} = max(L_{comp}, L_{w}, L_{ifm} \times G_{w}, L_{ofm} \times G_{w})
\end{equation}

\begin{figure}[t]
    \centering
    \includegraphics[width=0.5\textwidth]{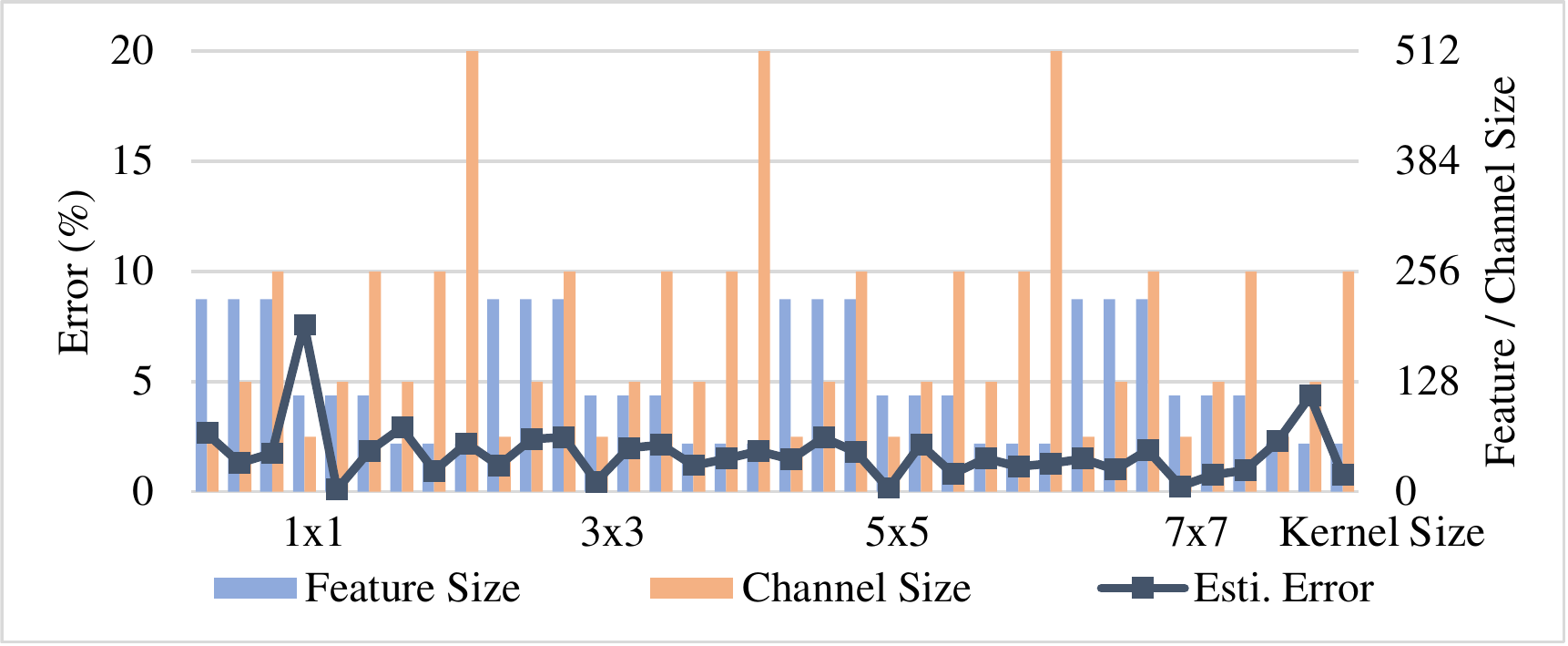}
    \vspace{-16pt}
    \caption{Performance estimation errors of the generic structure when mapping 36 cases of CONV layers, covering various feature map sizes (56, 112, 224), channel sizes (64, 128, 256, 512), and kernel sizes (1, 3, 5, 7).}
    \vspace{-12pt}
    \label{fig:esti_error_acc}
\end{figure}

On top of the estimation model, DNNExplorer adopts the DSE methodology from \cite{ye2020hybrid} to search for the optimized configuration of the generic architecture. As shown in Algorithm \ref{alg:gen_dse},  hardware parameters and data reuse strategies are configured according to given resource budgets.
To evaluate the analytical models of the generic structure, we include 36 cases of CONV layers with different channels, feature maps, and kernel configurations as benchmarks. We compare the estimated performance with the measured board-level performance using a Xilinx VU9P FPGA. Results are shown in Figure \ref{fig:esti_error_acc}, where only a 2.17\% average error is observed across all 36 cases, which guarantees actuate accelerator modeling.


%% file: sections/05-exploration.tex
\section{DNNExplorer for accelerator exploration}
\label{sec:exploration}


\subsection{Drawbacks of paradigm 1 and 2}

\begin{figure}[t]
  \centering
  \includegraphics[width=1.05\columnwidth]{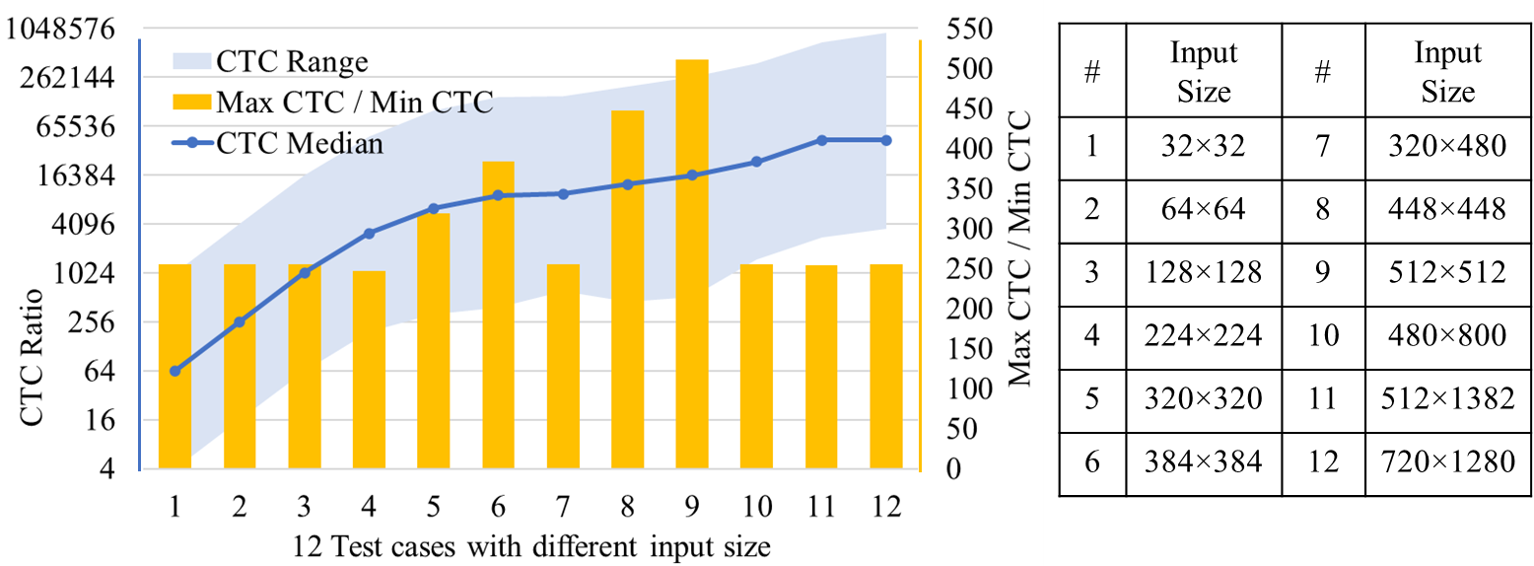}
   \caption{CTC distribution in VGG-16 models (without FC layers) regarding 12 input resolution cases. Inputs are RGB images with 3 depth channels and height and width listed as input size.}
   \vspace{-6pt}
   \label{fig:inputsize_ctc}
\end{figure}

By applying the optimization strategies mentioned in Section \ref{sec:accelerator_optim}, we can fully exploit the potential of accelerators following both pipeline and generic architecture under the predefined hardware constraints. 
However, there are still challenges preventing further performance improvements. It is hard for the generic accelerator to accommodate layers with various computation and memory demands using a generic compute unit, even though the layers from the same DNN model.
To indicate the various arithmetic intensity, we present the computation-to-communication (CTC) ratios of layers from 12 DNN models in Figure \ref{fig:inputsize_ctc}. The medians of CTC show an upward trend along with higher input resolutions. From 32$\times$32 to 512$\times$512 inputs, CTC medians rapidly increase by nearly 256 times following the blue curve, which implies significantly different computation and memory-access patterns. 
If using a generic compute unit to process these layers, we have to accept sub-optimal solutions because of the lack of architecture specificity.

To obtain more intuitive data, we use DSP efficiency (defined as Equation \ref{eq:dsp_eff}) to evaluate whether an accelerator is working efficiently when prototyping in FPGA. $GOP/s$ is a throughput performance metrics meaning Giga operations per second, while $\alpha$ represents the number of multiply-accumulate (MAC) operations handled by one DSP in one clock cycle, i.e.,  $\alpha=2$ for 16-bit and $\alpha=4$ for 8-bit inputs. Higher DSP efficiency means better utilization of the available computation resources.

\begin{equation}
\small
\label{eq:dsp_eff}
    EFFI_{DSP} = \frac{GOP/s}{\alpha\times DSP_{allocated}\times FREQ}
\end{equation}

\begin{figure}[t]
  \centering
  \vspace{-2pt}
  \includegraphics[width=1.05\columnwidth]{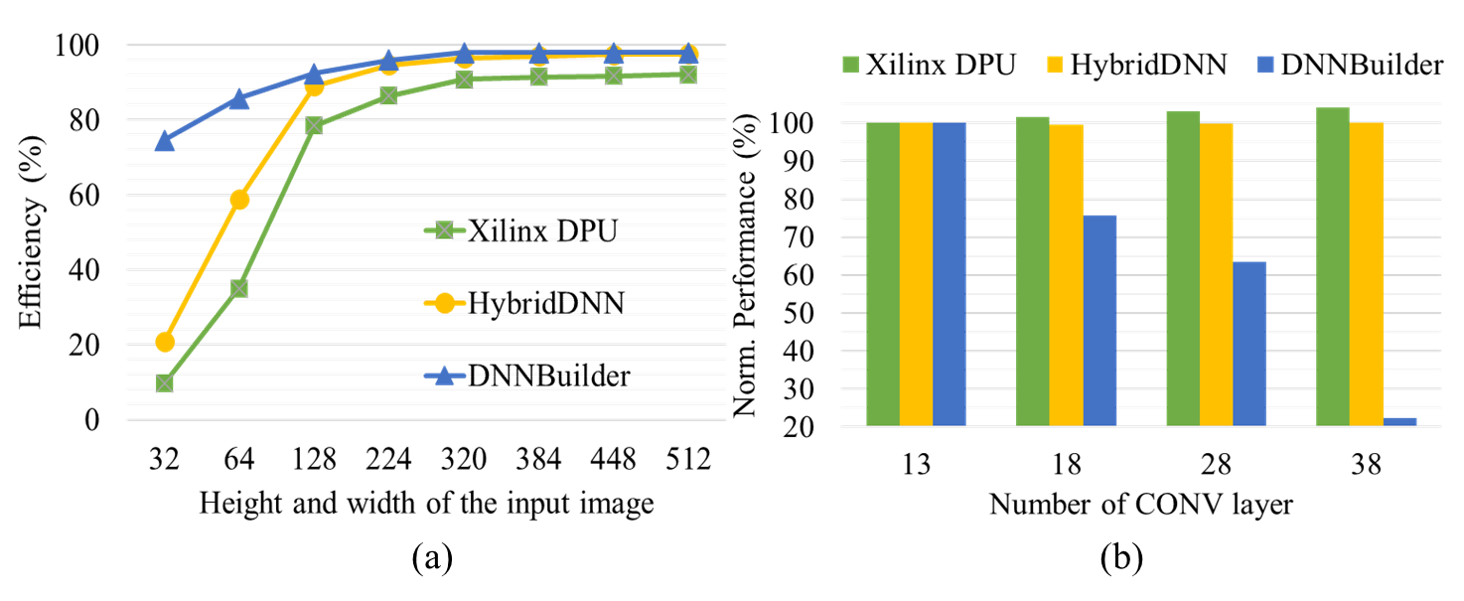}
  \vspace{-16pt}
   \caption{(a) DSP efficiency when running VGG16 with increasing input sizes in three representative FPGA-based DNN accelerators (batch size $=$ 1); (b) Normalized throughput performance in three accelerators when running VGG-like DNNs with 3$\times$224$\times$224 inputs and 13$\sim$38 CONV layers. (The performance of each accelerator is normalized to their baseline cases running the 13-layer DNN.)}
   \vspace{-10pt}
   \label{fig:inputsize_depth}
\end{figure}

We select three recently published DNN accelerators: Xilinx DPU \cite{xilinx_dpu}, HybridDNN \cite{ye2020hybrid}, and DNNBuilder \cite{zhang2018dnnbuilder} for comparison and present their DSP efficiency and throughput performance after benchmarking. In Figure \ref{fig:inputsize_depth} (a), both Xilinx DPU and HybridDNN (representing the generic architecture) suffer lower DSP efficiency (up to 64.9\% and 53.7\% degradation, respectively) compared to dedicated designs from DNNBuilder \cite{zhang2018dnnbuilder} (representing the pipeline architecture). 
By following the layer-based pipeline architecture, accelerators can adopt dedicated pipeline stages according to layers' inherent characteristics (e.g., arithmetic intensity, computation, and memory demands). It also enables more fine-grained hardware configurations and more in-depth design space exploration to address the low DSP efficiency issue. 
However, its scalability may be easily restricted by the number of DNN layers that it can support. A deeper DNN means more pipeline stages and fewer resources for each layer. In this case, performance degradation is expected as shown in Figure \ref{fig:inputsize_depth} (b), where accelerators need to handle 4 VGG-like DNNs with 13$\sim$38 CONV layers. The performance of DNNBuilder decreases 77.8\%  on a 38-layer model compared to the shallower network with 13 CONV layers. In contrast, generic accelerators maintain stable performance.

\subsection{The proposed architecture paradigm}

To overcome these challenges, we introduce a new paradigm to capture the essences of both existing paradigms but address their disadvantages. It is described as paradigm 3 in Figure \ref{fig:dnnexplorer_flow}.
For the first part, we implement a pipeline accelerator to generate dedicated layer stages for the first $SP$ layers. It helps guarantee higher DSP efficiency and more fine-grained resource allocation. For the second part, we adopt a generic architecture for the rest of DNN layers to support deeper DNNs with given resources. 
The reason for such a combination comes from a common phenomenon during DNN inference, where the first half of DNN layers (close to the input layer) contains more variances of arithmetic intensity compared to the second half, which easily causes low DSP efficiency when following the generic paradigm. 
In our proposed design, this problem can be properly resolved by the layer-dedicated pipeline architecture. By considering the second half, where layers have fewer variances, a generic structure can be adopted to improve the design scalability.
The proposed design is not a simple concatenation of the two existing paradigms,
as the fusion of two heterogeneous structures can directly cause an exponential increase of the design space and easily lead to tedious explorations and sub-optimal solutions. 
Therefore, we propose an automatic architecture exploration to deliver optimized architecture configurations for accelerators following the novel paradigm.

\subsection{Architecture exploration}

\subsubsection{Design space of the novel paradigm}
With the new accelerator paradigm, we are allowed to explore significantly larger design space and perform a more fine-grained hardware configuration. We define a dynamic design space regarding all possible accelerator design combinations in Table \ref{tab:design_space}. Split-point (SP) defines the task partitioning between the pipeline and generic structure. With more layers distributed to the pipeline structure, more stages are instantiated along with higher dimensions of $CPF_p$ and $KPF_p$. %
Given resource budgets $C_{max}, M_{max}, BW_{max}$ (corresponding to the available computation resources, on-chip memory resources, and external memory access bandwidth), DNNExplorer is designed to explore all design parameters and generate the best accelerators for targeted DNN applications.
By targeting accelerator prototyping on FPGA, $C_{max}$, $M_{max}$, and $BW_{max}$ represent the available DSPs, BRAMs, and external memory access bandwidth of the targeted FPGA.
To efficiently explore the design space, we adopt a divide and conquer strategy and propose a two-level automatic DSE engine to perform global and local optimization, which will be introduced in the next two subsections.

\begin{table}[t]
\footnotesize
\caption{Design space of the proposed paradigm}
\vspace{-6pt}
\label{tab:design_space}
\begin{center}
\newcommand{\tabincell}[2]{\begin{tabular}{@{}#1@{}}#2\end{tabular}}
\begin{tabular}{c|c|c}
\toprule
& Pipeline Structure & Generic Structure\\ \hline
\multirow{3}{*}{Parameters}  & \begin{tabular}[c]{@{}c@{}} \tabincell{c}{  $CPF_p = $ \{$CPF_1, CPF_2, \cdots CPF_i$\} \\ $KPF_p = $\{$KPF_1, KPF_2, \cdots KPF_i$\}  }\end{tabular}  & \begin{tabular}[c]{@{}c@{}} \tabincell{c}{  $CPF_g$ \\ $KPF_g$ \\$Buffer$-$allocation$\\$Dataflow$ }\end{tabular}
\\ \cline{2-3}
& \multicolumn{2}{c}{ $Split$-$point (SP)$ } \\ \cline{2-3}
& \multicolumn{2}{c}{ $Batch$ } \\ \hline

Budgets & \multicolumn{2}{c}{ $C_{max}, M_{max}, BW_{max}$ }   \\ 
\bottomrule
\end{tabular}
\end{center}
\end{table}

\subsubsection{A two-level DSE engine}
The first-level optimization goal is to determine task and resource partitioning schemes between the pipeline and generic structure. DNNExplorer introduces RAV (resource allocation vector defined by Equation \ref{eq:rav}) to provide guidelines for mapping the targeted DNN workloads and allocating resources to the customized accelerator following the proposed paradigm.
Layer $1 \sim$ $SP$ are mapped to a $SP$-stage pipeline structure with the resource budget $[DSP_p,BRAM_p,BW_p]$. Assuming the globally available resource is $[DSP, BRAM, BW]$, the rest of the layers are mapped into a generic reusable structure with the resource budget $[DSP-DSP_p,BRAM-BRAM_p,BW-BW_p]$. The pipeline and generic structure share the same batch size and working clock frequency. 

\begin{equation}
\small
RAV=[SP,Batch,DSP_p,BRAM_p,BW_p]
\label{eq:rav}
\end{equation}

To select the best RAV across a high-dimensional design space, we employ a particle swarm optimization (PSO) algorithm to discover the most suitable combination of task and resource partitioning schemes between the pipeline and generic structure. 
As shown in Algorithm \ref{alg:global}, each RAV is considered as a particle $P_i$, and all of them contribute to the swarm $\mathcal{M}$. We use throughput (GOP/s) as the fitness score to compare the quality of each design. To get the throughput estimation, we construct a fitness function using analytical models introduced in Section \ref{sec:accelerator_optim}.  By calling this function, the pipeline and generic structure optimization processes are individually executed to deliver the most suitable hardware configuration given the input RAV and the targeted workload. Once the configuration is ready, it is passed to analytical models for throughput (score) calculation. 
Based on the fitness score, we label the local best for particle $i$ across all iterations as $L_i$ and use $G$ to represent the global best particle. The search contains  $\mathcal{N}$ iterations, and in the $itr$-iteration, each particle's position is updated according to the current position and updated velocity $V_i$. $V_i$ is calculated based on the velocity to the local $V_{toLbest_i}$ and global $V_{toGbest_i}$ best position. 
In the update function, $rand()$ generates random numbers between 0 to 1 and we also include the adjustable parameters as inertia weight, $w$, acceleration constants, $c_1$ and $c_2$ to fine-tune the search process. Eventually, the best particle is selected, which is the best RAV indicating the optimal task partitioning and resource allocation. 

\begin{algorithm}[t!]
\renewcommand\baselinestretch{0.8}\selectfont
\footnotesize
\caption{The DSE algorithm}
\begin{algorithmic}[1]
\State Initialize RAV with $\mathcal{M}$ Population: $P(\mathcal{M})$
\State Initialize iteration number: $\mathcal{N}$
\State Initialize HW boundary: $SP_{max}$, $DSP_{max}$, $BRAM_{max}$, $BW_{max}$
\State Evaluate each RAV: $Fit_i= FitnessScore(P_i)$, where $i \in \mathcal{M}$
\State Keep the local best for each RAV: $L_i = Fit_i$, and the global best: $G=max(L_i)$
\State \textbf{While} $itr < \mathcal{N}$:
\State \quad \textbf{For} each $P_i$ in $\mathcal{M}$:
\State \quad \quad Get local and global velocity: $V_{toLbest_i}$, $V_{toGbest_i}$ 
\State \quad \quad Update velocity: $V_i = w \cdot V_i + c_1 \cdot rand() \cdot V_{toLbest_i} + c_2 \cdot rand() \cdot V_{toGbest_i} $
\State \quad \quad Update RAV: $P_i = UpdatePos(P_i,V_i)$
\State \quad \quad Evaluate updated RAV: $Fit_i = FitnessScore(P_i) $
\State \quad \quad Update local best: $L_i$
\State \quad Update global best: $G$
\State Output the best RAV
\end{algorithmic}
\label{alg:global}
\end{algorithm}

The second-level optimization is for individual optimization. 
Once the RAV is updated,  pipeline and generic optimization strategies are launched to search for the best hardware configurations for both pipeline and generic structure (e.g., $CPF_p$, $KPF_p$, $CPF_g$, $KPF_g$, etc.) given the constraint of RAV.
For the pipeline architecture, we follow strategies in Algorithm \ref{alg:pip_comp} and \ref{alg:pip_mem}. $CPF$ and $KPF$ are calculated for each pipeline stage based on workload, arithmetic intensity, and given resource budget defined by the RAV. 
Then in Algorithm \ref{alg:gen_dse}, DNNExplorer starts optimizing the generic structure to balance the pipeline throughput performance.

%% file: sections/06-results.tex
\section{Experimental results}

\begin{figure}[t]
    \centering
    \includegraphics[width=0.45\textwidth]{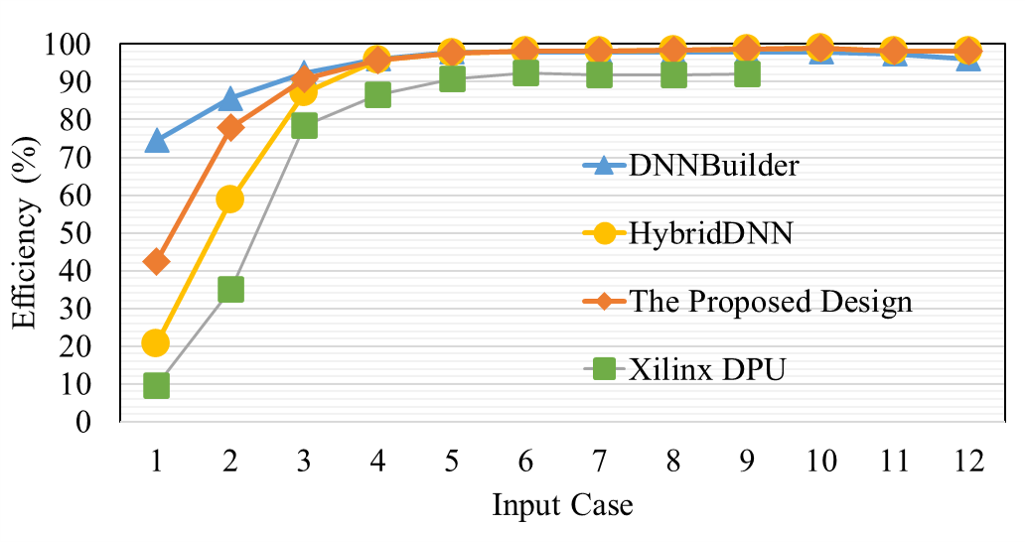}
    \vspace{-10pt}
    \caption{DSP efficiency comparison when running VGG16 (batch size = 1) with 12 different input sizes.}
    \label{fig:inputVSeff}
    \vspace{-6pt}
\end{figure}


\begin{figure}
    \centering
    \includegraphics[width=0.5\textwidth]{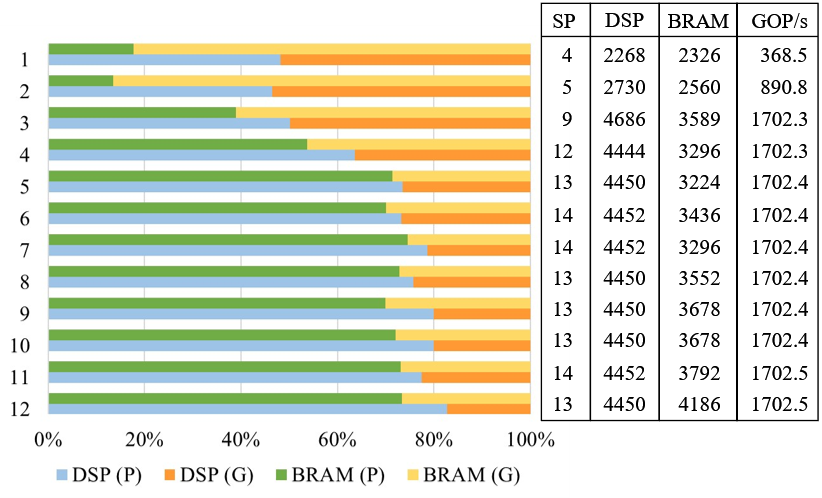}
    \vspace{-20pt}
    \caption{(left) Resource distribution between pipeline (P) and generic structure (G) in the proposed paradigm 3 when targeting VGG16 (batch size = 1) with 12 different input sizes. (right) Split-point (SP), utilization, and throughput of the generated accelerators.}
    \vspace{-4pt}
    \label{fig:inputsize_exp}
\end{figure}

\begin{figure}
    \centering
    \includegraphics[width=0.43\textwidth]{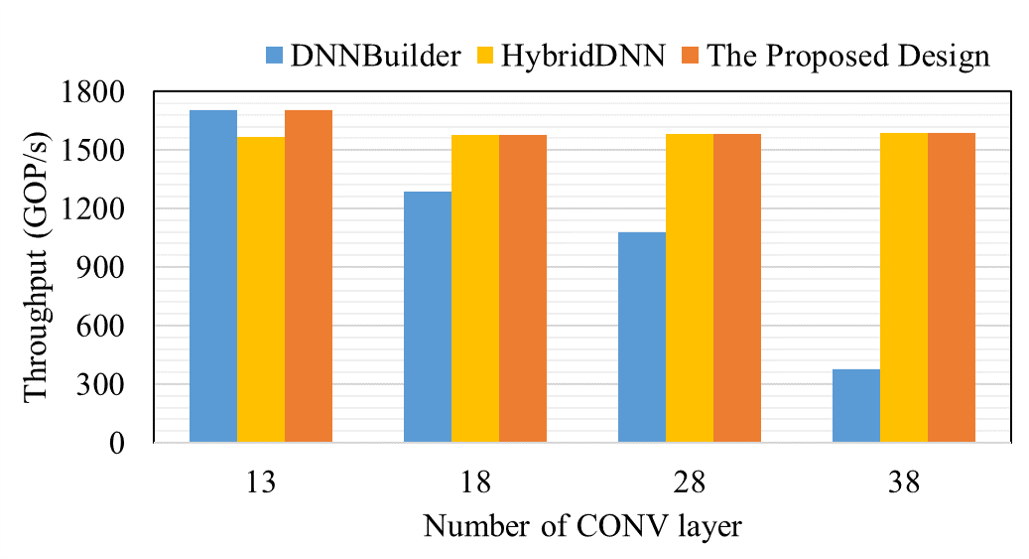}
    \vspace{-8pt}
    \caption{Throughput comparison when running deeper DNNs with the same $3\times224\times224$ input.}
    \label{fig:depthVSgops}
    \vspace{-6pt}
\end{figure}

In this section, we demonstrate DNNExplorer for customized hardware accelerator benchmarking and exploration. The proposed tool is connected to a PyTorch framework and takes the PyTorch-developed DNNs as inputs. We select popular DNNs from the torchvision library, including models in the VGG family, the ResNet family, and the AlexNet. For the targeted hardware platforms, we choose two FPGAs as Xilinx KU115 and Xilinx ZC706, to cover different scenarios for cloud and edge applications, respectively.

\begin{figure*}[t]
    \centering
    \vspace{-12pt}
    \includegraphics[width=1\textwidth]{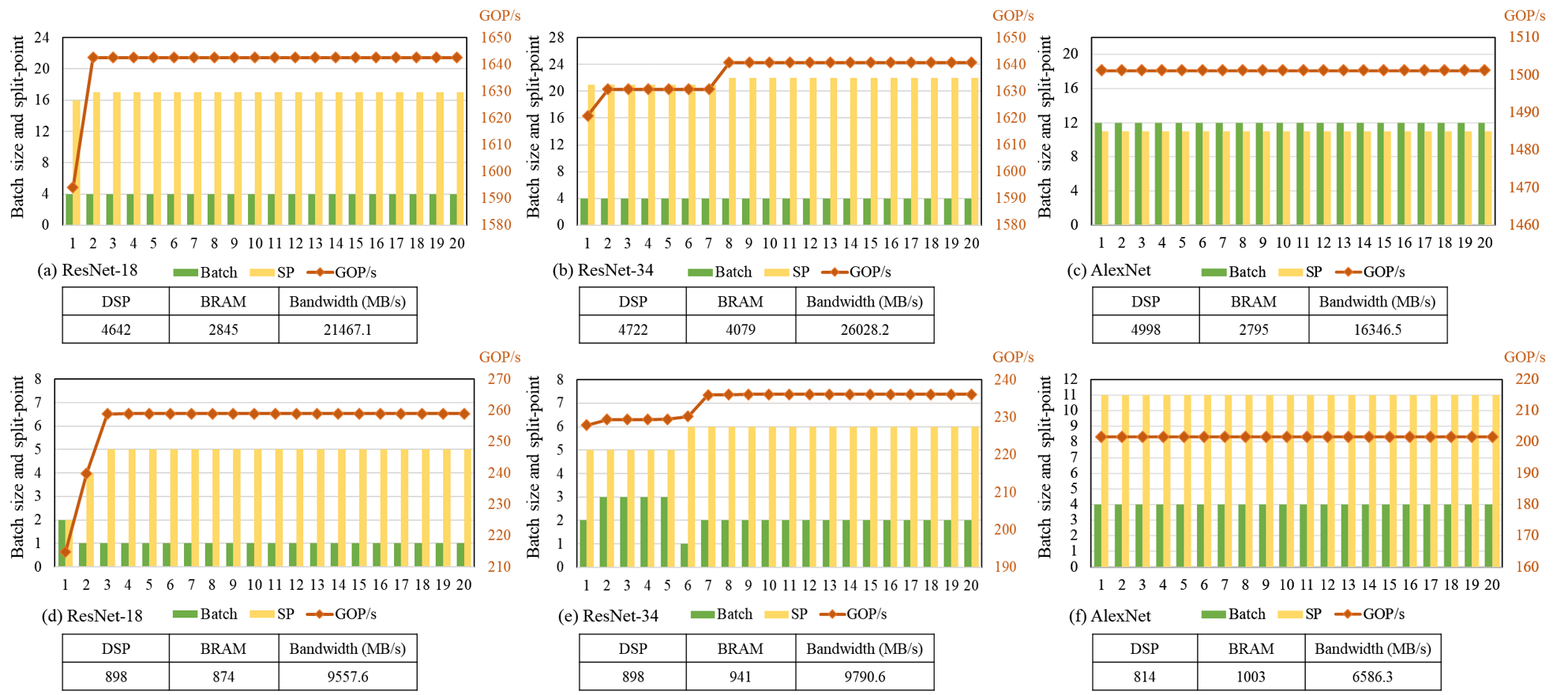}
    \vspace{-22pt}
    \caption{Architecture exploration for accelerators adopting paradigm 3 for ResNet-18/-34 and AlexNet. Designs in (a)$\sim$(c) adopt the resource budget of a KU115 FPGA while designs in (d)$\sim$(f) are targeting a ZC706 FPGA. Resource utilization is shown in the table below each sub-graph.}
    \vspace{-8pt}
    \label{fig:search_exp}
\end{figure*}

\subsection{DSP efficiency}
\label{sec:dsp_eff_comp}
We use DSP efficiency to evaluate whether accelerators maximize the usage of allocated computation resources for FPGA implementation. We adopt DNNExplorer to evaluate all three supported accelerator paradigms mentioned in Figure \ref{fig:dnnexplorer_flow}: the pipeline architecture from \cite{zhang2018dnnbuilder} (paradigm 1), the generic architecture HybridDNN \cite{ye2020hybrid} (paradigm 2), and the proposed novel architecture (paradigm 3). All three accelerator designs are targeting the same set of DNNs: 12 VGG-16 (without the last three FC layers) models with different input sizes (which are listed in Figure \ref{fig:inputsize_ctc} from \#1 to \#12) and the same KU115 FPGA for a fair comparison. We also quantize the DNN models to use 16-bit fixed-point data.
Benchmarking results are shown in Figure \ref{fig:inputVSeff}, where paradigm 1 achieves the highest DSP efficiency, especially for small-sized inputs (e.g., case 1 and 2) because of its dedicated pipeline stage design. The proposed paradigm 3 is slightly behind when targeting small inputs but soon reaching the same efficiency level (>95\%) after case 3. Compared to the generic accelerator design, paradigm 3 can deliver 2.0$\times$ and 1.3$\times$ higher efficiency for case 1 and case 2, respectively.
We also compare to Xilinx DPU \cite{xilinx_dpu}, a commercial DNN accelerator IP, for running the first nine cases on a ZCU102 FPGA (as the last three inputs are not supported yet). The proposed paradigm 3 can achieve an average 1.6$\times$ higher DSP efficiency, peaking at 4.4$\times$ for case 1. As the input size increases, the efficiency gap decreases (<10\% after case 5).

\subsection{Throughput performance}
We present the throughput performance of the accelerators following paradigm 3 in Figure \ref{fig:inputsize_exp}. These accelerators target the same DNNs (mentioned in Subsection \ref{sec:dsp_eff_comp}) and the same KU115 FPGA with 200MHz working frequency.  
We restrict the batch size to one and create a more demanding application scenario as accelerators can not always rely on larger batch sizes to reach higher throughput. 
The first two cases can not provide enough workloads (due to the smaller input size), so accelerators failed to reach their peak performance.
To meet resource constraints and maximize performance, the proposed two-level DSE engine generates task partitioning and resource distribution, where the SP indicates the split-point for partitioning the targeted DNN to the pipeline and generic structure and the stacked bar chart shows the resource allocation for DSP and BRAM resources. From the results, the proposed DSE is likely to allocate more tasks and resources to the pipeline structure when increasing the DNN input size from case 1 to case 12.

\subsection{Scalability}
To evaluate all three paradigms' scalability, we prepare four deeper DNNs, with 13, 18, 28, and 38 CONV layers. The 13-layer one comes directly from VGG16 by removing the FC layers. Since VGG is composed of five CONV groups, where each group has the same CONV configurations (e.g., number of CONV kernels), we add one CONV layer to each group (maintaining the same configurations) and get the 18-layer (13+5) model. Similarly, we add three and five CONV layers to each part for the 28- and 38-layer model, respectively. In Figure \ref{fig:depthVSgops}, we can clearly observe the drawback of using paradigm 1, which is hard to handle a deeper network. By contrast, accelerators following paradigm 2 and 3 can maintain peak performance despite targeting deeper networks. Compared to paradigm 1, the proposed paradigm 3 can deliver 4.2$\times$ higher performance for accelerating a 38-layer VGG-like DNN on the same FPGA platform.

\subsection{Architecture exploration}
To better understand the architecture exploration feature provided by DNNExplorer, we extend the experiment for customized accelerator design following paradigm 3. We lift the batch size restriction and capture the hardware configuration changes while applying the proposed two-level DSE. We target three DNNs (ResNet-18/-34 and AlexNet from torchvision) and two hardware platforms (KU115 and ZC706) for this experiment, and results are shown in Figure \ref{fig:search_exp}. 
Since we set the search iteration number to 20, each sub-figure has 20 pairs of green and yellow bars, indicating the batch size and split-point configuration. We also present the throughput of the best results for every iteration along the red curve. 
Resource utilization of the best accelerator after exploration is shown below the corresponding sub-graph.
The proposed two-level DSE engine can soon search for the optimized hardware configuration with the throughput performance reaching the peak during the first ten iterations for all six cases. Shown in Figure \ref{fig:search_exp} (a)$\sim$(c), the customized accelerators deliver 1642.6, 1640.6, and 1501.2 GOP/s using KU115 after exploration, while in Figure \ref{fig:search_exp} (d)$\sim$(f), accelerators deliver 258.9, 236.1, and 201.6 GOP/s using ZC706.


To accommodate the rapid development of DNN, DNNExplorer follows the modular design strategy, which can be extended to support more emerging layers by adding corresponding layer modules. With these extended models, newly supported layers can be parsed in step 1 of the proposed flow (Figure \ref{fig:dnnexplorer_flow}) to extract layer-wise information. Still, DNNExplorer has not yet supported some new network architectures such as the Transformer \cite{vaswani2017attention} as it requires additional support of unique layer modules for improving the accelerator architecture paradigm. To support more network architectures will be our future work.

%% file: sections/09-conclusion.tex
\vspace{-5pt}
\section{Conclusion}
\vspace{-2pt}

This paper presented DNNExplorer, an automation tool for benchmarking and exploring customized DNN hardware accelerators. 
Novel technologies included the direct support of popular machine learning frameworks for easy access to the existing or customized DNNs; highly accurate analytical models for effective hardware accelerator benchmarking; a novel accelerator paradigm to overcome drawbacks of the existing paradigms; and a two-level DSE engine for accelerator exploration to deliver optimized accelerators by considering both targeted DNN workloads and given resource budgets.
With the above technologies, DNNExplorer can provide customized accelerator benchmarking and perform architecture exploration to deliver optimized solutions when targeting emerging AI applications. It can also offer DNN and accelerator design insights to enable more efficient AI deployments at the earliest design stage.

\section*{Acknowledgment}
\vspace{-2pt}
We want to thank Junsong Wang, Yonghua Lin, Jinjun Xiong, and Wen-mei Hwu, who contributed to an early version of the work published in ICCAD'20. This work is partly supported by IBM-Illinois Center
for Cognitive Computing System Research (C3SR) and a Google PhD fellowship to Xiaofan Zhang.

\vspace{-4pt}